\documentclass[12pt]{article}

\usepackage{extsizes}
\usepackage[super,sort&compress,comma]{natbib}
\usepackage[version=4]{mhchem}

\usepackage[left=2.5cm, right=2.5cm, top=2.5cm, bottom=2.5cm]{geometry}

\usepackage{balance}
\usepackage{mathptmx}
\usepackage{sectsty}
\usepackage{graphicx}
\usepackage{lastpage}

\usepackage[format=plain, justification=justified, singlelinecheck=false,  font={stretch=1.125,small,sf}, labelfont=bf, labelsep=space]{caption}

\usepackage{float}
\usepackage{fancyhdr}
\usepackage{fnpos}
\usepackage[english]{babel}
\usepackage{array}
\usepackage{booktabs}
\usepackage{multirow}
\usepackage{xcolor}
\usepackage{hyperref}

\usepackage{subcaption}

\addto{\captionsenglish}{%
    
}

\begin{document}
\title{Designing Geared Molecular Rotors in\\
Flexible Metal-Organic Frameworks}

\author{%
Konstantin Stracke$^{1}$ and Jack D. Evans$^{1,*}$\\[0.75em]
\small $^{1}$School of Physics, Chemistry and Earth Sciences,\\
\small Adelaide University, City East Campus,\\
\small Adelaide, South Australia 5005, Australia\\[0.5em]
\small $^{*}$Corresponding author:
\href{mailto:j.evans@adelaide.edu.au}{j.evans@adelaide.edu.au}
}
\date{}

\maketitle

\begin{abstract}
Translating macroscopic mechanical concepts such as gears to the molecular scale offers a route to precise control over nanoscale motion, inspired by the efficiency of biological machines. In this work, we explore two-dimensional interlocked linker rotations in the flexible metal-organic framework NU-2002. The volume-dependent free energy landscape is mapped using bias-deletion metadynamics to identify six characteristic phases, ranging from very closed to very open pore. We then perform unbiased molecular dynamics simulations at each phase to track rotation events and evaluate co-rotation statistics, calculate free energy barriers, and measure pairwise radial force transfer between neighbouring linkers. We show that the free energy barrier acts as a rotational-speed filter, where rare rotational transitions lead to higher speeds on average. Across the volume profile, we classify four distinct geared regimes: sterically restricted rotations at small volumes (Jammed Gear), rotational coupling extending to third neighbour linkers (First Gear), in-plane partnered linker coupling (Second Gear), and uncoupled, independent rotation at large volumes (Neutral Gear). Together, these results establish framework breathing as a molecular gearbox that switches both the speed and collective coupling of linker rotation. Framework flexibility therefore offers a handle for programming collective motion in solid-state molecular machines.
\end{abstract}

\section*{Introduction}
Molecular machines derive their function from the controlled motion of molecular components, yet translating such motion into extended solids remains a major challenge.\cite{feng2022controlling, krause2020towards}
Metal-organic frameworks (MOFs) provide a particularly versatile platform for addressing this challenge because their inorganic nodes and organic linkers can be selected and functionalised with high precision.\cite{furukawa2013chemistry}
In flexible MOFs, this tunability is complemented by large and reversible changes in pore geometry.\cite{schneemann2014flexible}
Importantly, the organic linkers are not necessarily static components of the framework.
The motion of framework components can manifest as molecular vibration or rotation\cite{gonzalez2019rotational} and can even contribute to macroscopic responses such as thermal expansion\cite{evans2019assessing}, which can switch between positive and negative in the different phases of a flexible MOF\cite{stracke2025investigating}.
Interactions between neighbouring linkers can correlate these motions, producing collective rotational behaviour rather than independent molecular rotation.
This coupling should be particularly sensitive to framework flexibility because changes in pore volume alter the distances, orientations, and non-bonded interactions between neighbouring linkers.

\begin{figure}
\begin{center}
\includegraphics[width=0.5\textwidth]{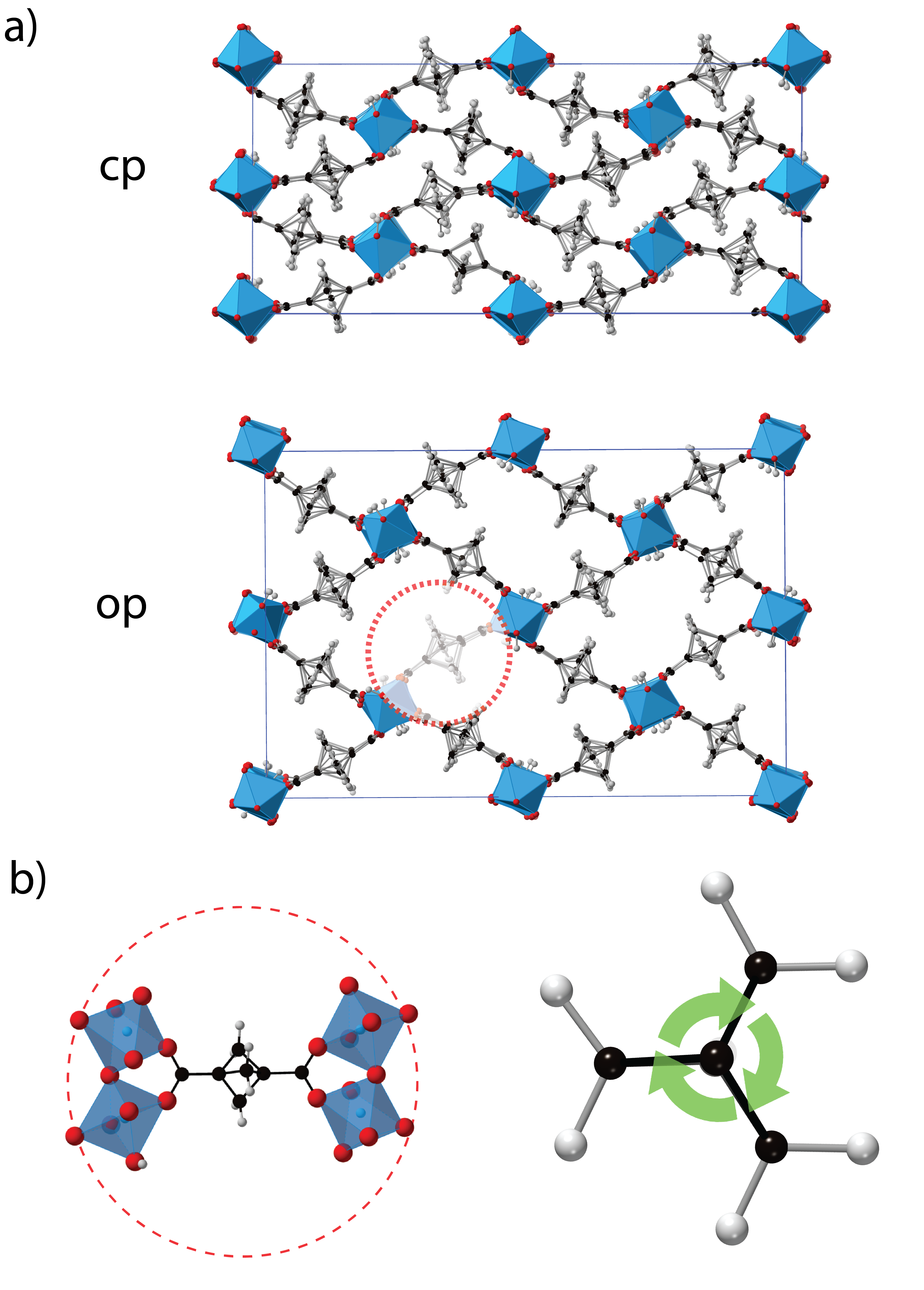}
\caption{(a) The MOF {NU-2002} in the closed pore (\emph{cp}) and open pore (\emph{op}) phases. (b) An isolated linker viewed from two directions, with the rotation about the linker axis indicated.}
\label{intro:gears}
\end{center}
\end{figure}

MIL-53 and related wine-rack frameworks provide prototypical examples of large amplitude framework flexibility.\cite{loiseau2004rationale}
Previous studies have demonstrated coupled motion of phenylene rotors with substituents in these frameworks, including the emergence of long range rotational correlations.\cite{gonzalez2021emergence,mula2024long}
However, these studies primarily consider phenylene based or functionalised rotors and do not systematically examine how changes in the framework volume modify the rotational dynamics and coupling between the linkers themselves.
Moreover, the phenylene linkers in MIL-53 undergo flips rather than rapid, quasi-continuous rotation.\cite{kolokolov2014guest}
NU-2002 (Figure~\ref{intro:gears}) is a promising system that combines an aluminium based flexible wine-rack-related framework with three-dimensional bicyclo[1.1.1]pentane-1,3-dicarboxylate linkers.\cite{smoljan2023engineering}
Bicyclo[1.1.1]pentane-based units are capable of exceptionally rapid molecular rotation, making NU-2002 suitable for examining the coupling between intrinsic linker rotation and framework breathing.\cite{prando2020molecular, perego2020fast}

Here, we determine how pore contraction and expansion regulate linker rotation in NU-2002.
Using bias-deletion metadynamics, we first map the volume-dependent free energy landscape of the framework and identify six characteristic phases spanning the breathing pathway.
Unbiased molecular dynamics simulations at each phase then resolve the rotational free energy barriers, rotation speeds, co-rotation statistics of neighbouring linkers, and the radial forces transferred between them.
We show that the framework acts as a low-speed rotational filter and that breathing switches the linkers between four distinct regimes, from jammed, through two geared regimes, to free and independent rotation.
Framework flexibility therefore emerges as a handle for controlling both the speed and the collective coupling of molecular rotors in extended solids.

\section*{Computational Details}
All simulations were conducted with the YAFF (Yet Another Force Field, version 1.6.0.post29) program package.\cite{verstraelen2022yaff}
For faster calculation of the interatomic forces, the LAMMPS (version 31 August 2021) implementation of parallel algorithms was used for the volume biased simulations.\cite{plimpton1995fast}

The NU-2002 simulation system consisted of a supercell containing 16 unit cells ($2\times2\times4$) with 64 linkers.
The temperature for all simulations was kept at 300~K employing a single Nos\'{e}--Hoover chain of length three and a relaxation time of 0.1~ps.\cite{nose1984molecular, nose1984unified, hoover1985canonical, martyna1992nose}
The force field for NU-2002 was generated by incrementally combining the metal node parameters from the {MIL-53(Al)} force field\cite{rogge2019modeling} with newly fitted parameters of the linker to DFT calculations by following the QuickFF procedure.\cite{vanduyfhuys2018extension}
Given the particular interest in rotation, the parameters for the linker rotation were refitted to DFT rotational barriers.
A detailed description of the force field fit is in our previous work.\cite{stracke2026bias}
Electrostatic interactions were treated using Ewald summation with fixed charges, a real-space cutoff of $12~\text{\AA}$, $\alpha_{\mathrm{scale}}=3.2$, $g_{\mathrm{cut,scale}}=1.5$, and no smoothing of the real-space contribution, while analytical tail corrections were applied to the non-electrostatic pair interactions.

The volume of {NU-2002} was biased for repeated pathway sampling with bias-deletion metadynamics (BDMTD).\cite{stracke2026bias}
A bias width and height of 175~Å$^3$ and 10~kJ~mol$^{-1}$ per unit cell were used. The bias was updated every 10,000 steps (5~ps) for a total of 1,500~biases or 5 repetitions of 300 biases leading to 15,000,000 MD steps, 7.5~ns.
The different pathways for each individual repetition are shown in Figure~S1, and the convergence for the total run in Figure~S2.
Representative structures were extracted at each characteristic volume; for each feature of the free energy profile, the sampled structure with the closest volume was chosen.
For every structure, three independent constant-volume molecular dynamics simulations were performed at 300~K using three different random seeds.
Cell-shape fluctuations were permitted at fixed volume using the MTK formalism with a relaxation time of 1~ps.\cite{martyna1994constant, rogge2019modeling}
Each simulation was propagated for 250~ps using a timestep of 0.5~fs.
The first 50~ps were discarded as equilibration, and atomic coordinates were recorded every 100 MD steps, corresponding to a sampling interval of 50~fs.
The calculation of the transferred radial force is detailed in Section~S4 of the Supplementary Information.

\section*{Results and Discussion}
\subsection*{Phases of NU-2002}
\begin{figure*}
\begin{center}
\includegraphics[width=\textwidth]{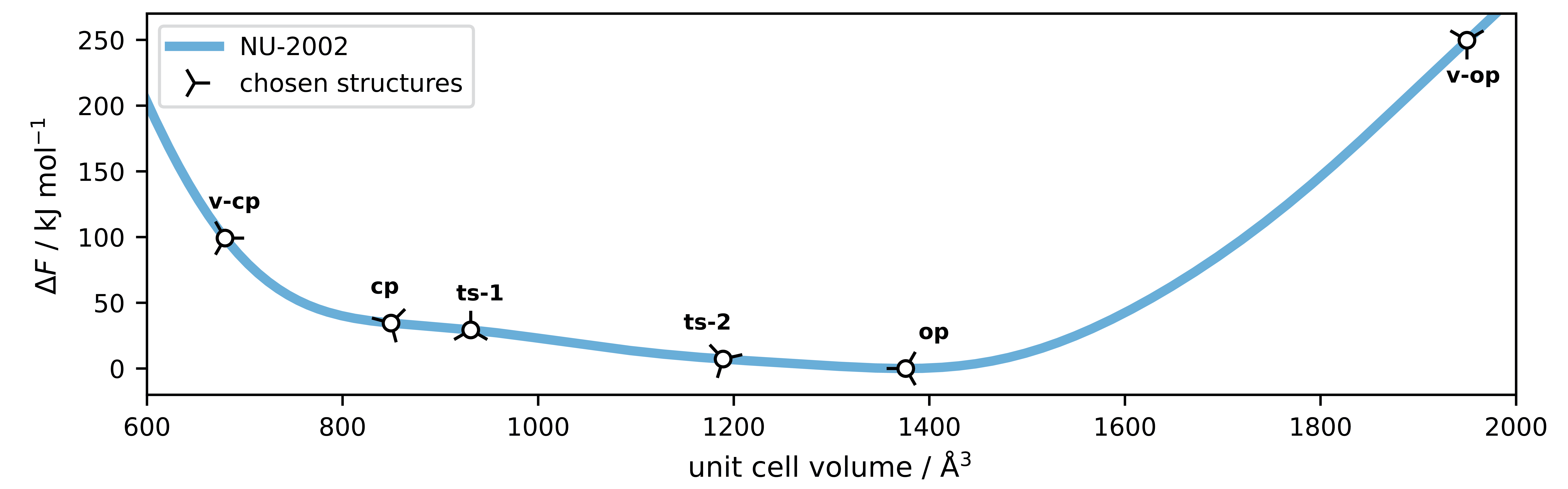}
\caption{The free energy profile of {NU-2002} is shown as a function of unit-cell volume. The configurations chosen for further analysis are highlighted.}
\label{result:gears_1}
\end{center}
\end{figure*}

Before the phase dependence of linker rotation can be examined, the different phases of {NU-2002} must be established.
Bias-deletion metadynamics repeatedly samples the free energy landscape, thus limiting artefacts from competing pathways by averaging over all sampled pathways (Figure~S1).\cite{stracke2026bias}
The resulting converged free energy profile is shown in Figure~\ref{result:gears_1}.
Two minima emerge, a shallow closed pore (\emph{cp}) phase near $850\ \text{\AA}^3$ and a broad open pore (\emph{op}) phase near $1389\ \text{\AA}^3$.
Experimentally, NU-2002 exhibits a faint \emph{cp} phase near $835\ \text{\AA}^3$ and an \emph{op} phase near $1125\ \text{\AA}^3$.
The volume discrepancy for the open pore structure likely stems from the wide free energy minimum. Consistent with the calculated free energy profile, the \emph{op} phase is thermodynamically more stable than the \emph{cp} phase.\cite{smoljan2025exceeding,smoljan2023engineering}

This free energy landscape provides the basis for examining how linker dynamics evolve across the MOF breathing pathway.
Representative configurations were selected at six characteristic volumes: very closed pore (\emph{v-cp}), closed pore (\emph{cp}), first transition state (\emph{ts-1}), second transition state (\emph{ts-2}), open pore (\emph{op}), and very open pore (\emph{v-op}).
Their positions along the free energy profile are indicated in Figure~\ref{result:gears_1}.
Together, these configurations span the breathing pathway and enable a systematic comparison of how linker rotations vary between phases.

\subsection*{Rotational Speed Filtering}
We next examine how pore volume affects the rotation speed of individual linkers.
For each selected phase volume, three independent molecular dynamics simulations of 250~ps were performed using different random seeds.
The dihedral angle of each linker was recorded throughout each MD trajectory. To identify rotation events, the signed angular displacement of each linker was accumulated over time (Figure~S4).
Displacements in opposite directions therefore cancel, excluding reversible back-and-forth motion.
A rotation event was registered when the accumulated displacement reached $\pm120^\circ$, corresponding to one rotational step of the threefold linker ($3\times120^\circ=360^\circ$, Figure~\ref{intro:gears}b).
The dihedral angle definitions employed here (o–ca–cb–cc) are shown in Figure~S3.
These identified rotation events were then used to determine the fraction of time spent rotating and the mean angular speed during rotation, $\langle|\omega|\rangle_{\mathrm{rot}}$~(Table~\ref{tab:data}).
The mean event duration was determined to be 23~frames (1.15~ps), and for comparison the rotational speeds over the entire trajectory were calculated using a matching 23~frame sliding window.
The mean rotational speeds, averaged over the complete trajectory ($\langle|\omega|\rangle$) and all linkers, are also reported in Table~\ref{tab:data}.

\begin{table*}
  \centering
\caption{Selected NU-2002 phases, their corresponding volumes, and linker rotation dynamics. The overall mean angular speed, $\boldsymbol{\langle|\omega|\rangle}$, is calculated over the complete trajectory. The rotating time fraction is the fraction of time classified as rotational motion and $\boldsymbol{\langle|\omega|\rangle_{\mathrm{rot}}}$ is the mean angular speed calculated only while the linkers are rotating. Uncertainties are reported as $\pm$ one standard deviation across independent seeds.}
  \label{tab:data}
  \begin{tabular}{@{\hskip 8pt} l c c c c @{\hskip 8pt}}
    \toprule
     \textbf{Phase}& \textbf{Volume} & $\boldsymbol{\langle|\omega|\rangle}$  & \textbf{Rotating time fraction} & $\boldsymbol{\langle|\omega|\rangle_{\mathrm{rot}}}$\\
  & \AA$^3$ & deg ps$^{-1}$ & \% &deg ps$^{-1}$ \\
    \midrule
    v-cp & 685 & 11.9  $\pm$ 0.2&  0.2 $\pm$  0.03 & 152.9 $\pm$ 5.1\\
    cp & 847   & 36.4  $\pm$ 0.2 & 12.0 $\pm$  0.14 & 123.4 $\pm$ 0.6 \\
    ts-1 & 925 & 38.5 $\pm$ 0.4  & 11.6 $\pm$  0.38 &137.8 $\pm$ 0.2 \\
    ts-2 & 1193 &  31.3  $\pm$ 0.3 & 14.0 $\pm$  0.46 & 157.6 $\pm$ 1.1 \\
    op & 1389   & 66.6   $\pm$ 0.2 & 51.7 $\pm$  1.40 & 128.3 $\pm$ 1.4\\
    v-op & 1959 & 73.8  $\pm$ 0.2 & 84.5 $\pm$  0.25 & 118.3 $\pm$ 2.7 \\
    \bottomrule
    \end{tabular}
\end{table*}

Table~\ref{tab:data} reveals two distinct effects of pore expansion on linker dynamics.
The overall mean angular speed, \(\langle|\omega|\rangle\), generally increases towards the expanded phases as the linkers spend an increasingly large fraction of the trajectory undergoing rotational motion.
By contrast, the mean angular speed during identified rotation events, $\langle|\omega|\rangle_{\mathrm{rot}}$, does not increase with volume but is instead lowest in the fully expanded \emph{v-op} phase.
Thus, expansion predominantly increases the accessibility of rotational motion rather than the speed of individual rotations; the restricted phases permit fewer rotations while preferentially excluding slower ones.
This behaviour is consistent with a sterically induced rotational filter, which suppresses attempts that lack the kinetic energy to overcome the barriers.
However, this trend includes an outlier, the \emph{ts-2} phase, which demonstrates that the filtering is governed not only by unit cell volume but by the rotational free energy landscape.

\begin{figure}
\begin{center}
\includegraphics{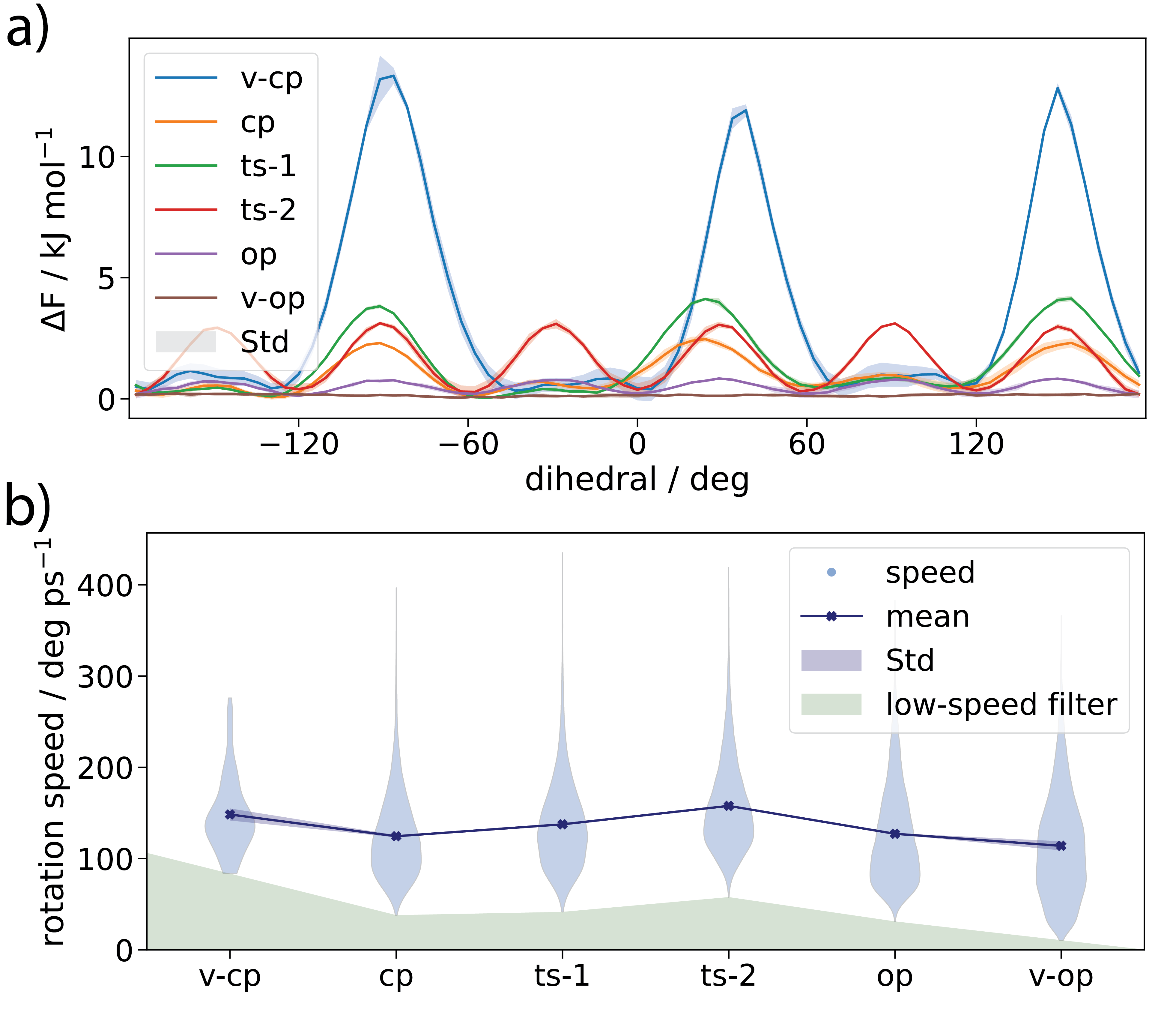}
\caption{(a) The linker rotational free energy profiles for each phase. (b) Violin plot of the linker rotation speeds, with the low-speed filter region highlighted in green. In both panels, the shaded regions indicate the standard deviation among the independent simulation seeds.}
\label{result:gears_2}
\end{center}
\end{figure}

To relate the observed rotation speeds to the rotational free energy barriers, the recorded dihedral angles were pooled across all linkers and trajectories to obtain the probability distribution $P(\phi)$ for each phase.
The corresponding Helmholtz free energy profile was subsequently calculated as
\begin{equation}
\Delta F=-RT\ln\left[\frac{P(\phi)}{P(\phi)_{\max}}\right],
\end{equation}
where $R$ is the molar gas constant, $T$ is the temperature, and $P(\phi)_{\max}$ is the maximum probability within each distribution.
The dihedral free energy profiles for the selected phases are shown in Figure~\ref{result:gears_2}a.
We find the rotational barriers decrease in the order \emph{v-cp}, \emph{ts-1}, \emph{ts-2}, \emph{cp}, \emph{op}, and \emph{v-op}.
Notably, the \emph{v-cp}, \emph{cp}, and \emph{ts-1} phases exhibit three barriers, whereas the remaining phases exhibit six.
This change in number of barriers reflects a transition from rotation governed by framework effects, particularly linker stacking, in the closed phases to rotation governed by the intrinsic torsional potential of the linker in the open phases.\cite{stracke2026bias}
More barriers of similar height appear to favour fast rotation, as seen for \emph{ts-2}, which shows the highest rotation speeds.
A more continuous rotational free energy landscape could therefore be exploited to design faster rotors.

For overall comparison the distributions of speeds across all identified rotation events were pooled for each phase and are shown as violin plots in Figure~\ref{result:gears_2}b.
The trend in Table~\ref{tab:data} is mirrored here: the lower mean rotation speed during rotation events, $\langle|\omega|\rangle_{\mathrm{rot}}$, in the \emph{v-op} phase arises from a population of slow rotations that is absent from the more confined phases.
The contracted framework therefore acts as a low-speed filter: the slower attempts are less likely to complete a $120^\circ$ transition, leaving predominantly faster events. By contrast, the low barriers in the \emph{v-op} phase permit both slow and fast rotations, increasing the number of events while lowering their mean speed.

\subsection*{Shifting Gears}
\begin{figure}
\begin{center}
\includegraphics{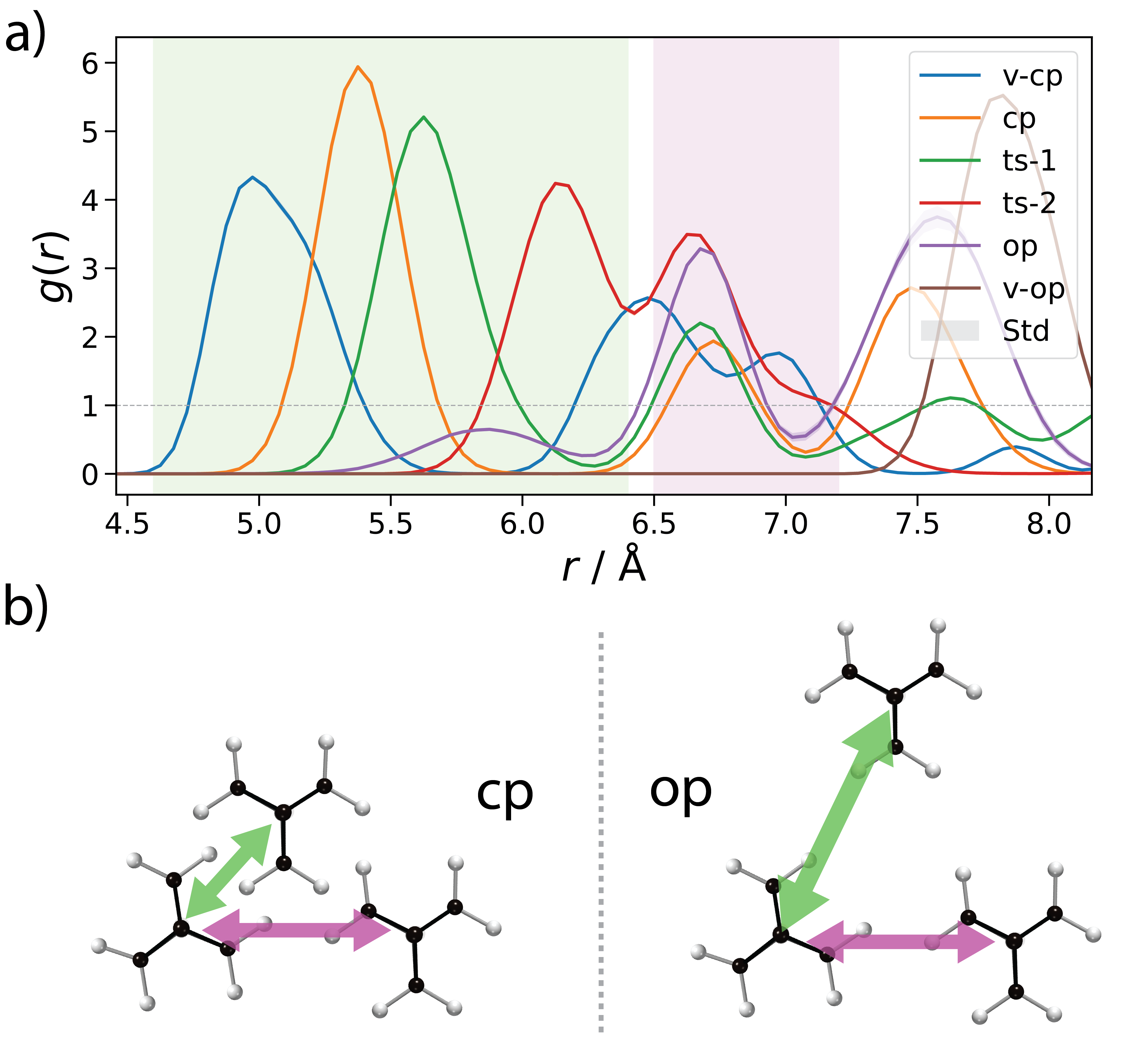}
\caption{(a) RDFs between linker centroids. The in-plane neighbour distance remains constant (purple shading), whereas the interlayer neighbour distance increases with volume (green shading). (b) Schematic highlighting the change in the closest neighbour from interlayer to in-plane with increasing volume.}
\label{result:gears_3}
\end{center}
\end{figure}

The key structural feature of {NU-2002} is its characteristic wine-rack architecture: the linkers are connected in-plane via the backbone, while hinge-like connections along the backbone couple neighbouring rows above and below (Figure~\ref{intro:gears}a).
The framework expands and contracts through hinge-like motion in the metal rows, causing adjacent layers to separate or approach like the folds of an accordion.
To quantify how this accordion-like motion changes the local linker environment, radial distribution functions between linker centroids were calculated, as shown in Figure~\ref{result:gears_3}a.
The \emph{v-cp}, \emph{cp}, \emph{ts-1}, and \emph{ts-2} phases exhibit a close neighbour peak below $6.5$~\AA, which shifts to larger distances as the framework volume increases.
All phases exhibit a neighbour peak between $6.5$ and $7.3$~\AA{} and this position remains unchanged with volume, except for \emph{v-op}.
These observations follow the described mechanism of expansion and contraction.
The closest neighbours in the contracted phases belong to the adjacent layers and move apart during expansion, whereas the peak between $6.5$ and $7.3$~\AA{} corresponds to the in-plane neighbours with distances that remain constant.
This mechanism is showcased in Figure~\ref{result:gears_3}b.

\begin{figure*}
\begin{center}
\includegraphics[width=\textwidth]{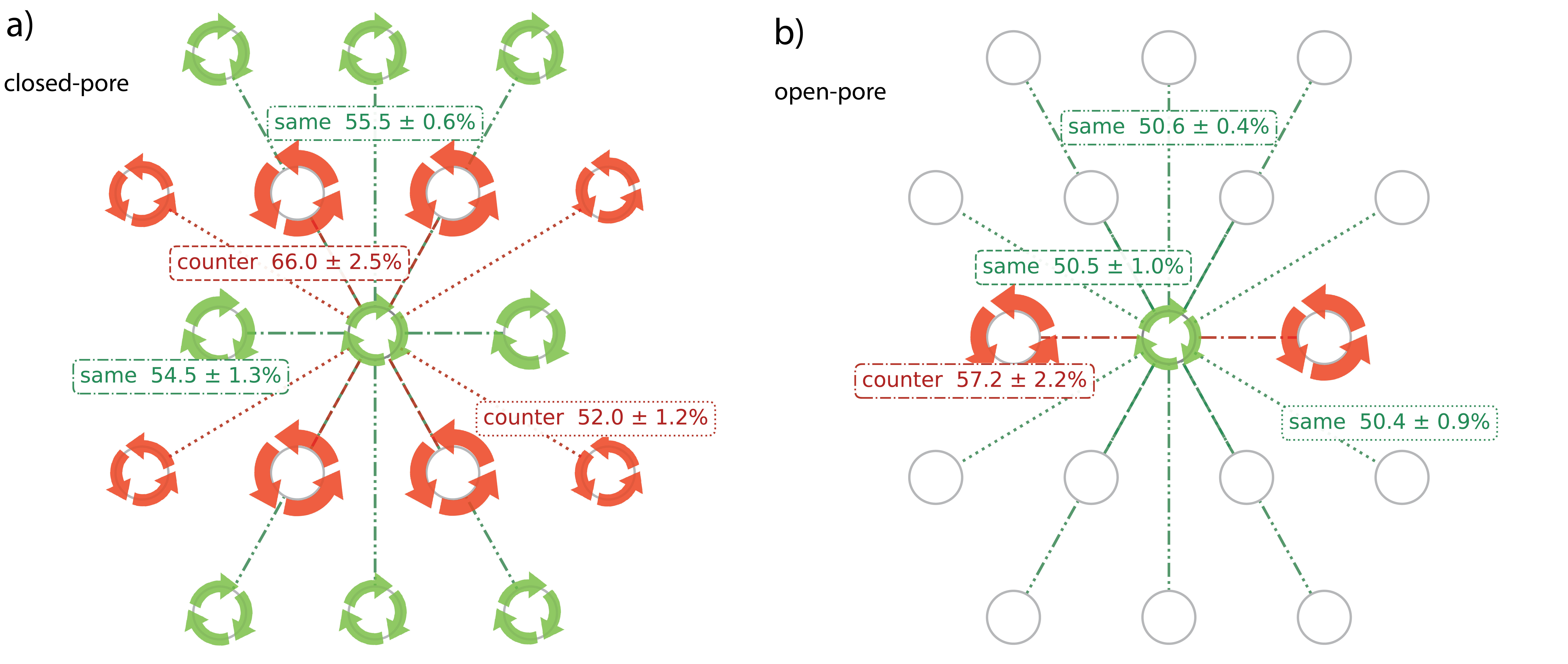}
\caption{Propagation of same-direction and counter-direction co-rotation through the linker neighbourhood in the (a) \emph{cp} and (b) \emph{op} phases, viewed along the linker rotational axis. Percentages are averaged over all linkers taken as the central linker, and uncertainties represent the standard deviation between seeds.}
\label{result:gears_4}
\end{center}
\end{figure*}

In contrast to the preceding analysis of individual rotation (see Rotational Speed Filtering), we also consider co-rotation events.
A co-rotation event is defined as an interval during which two linkers are simultaneously rotating, with a minimum overlap of 150~fs.
Each co-rotation event is classified as either same-direction rotation or counter-rotation.
To examine how far directional coupling extends through the framework, the co-rotation analysis is expanded to progressively more distant linkers in the surrounding neighbourhood, as illustrated in Figure~\ref{result:gears_4}.
Notably, each comparison is strictly pairwise.

In the most contracted phase, \emph{v-cp}, no co-rotation events are detected because rotations are exceptionally rare.
The confining arrangement of the diagonal neighbours blocks the rotation and co-rotation; this behaviour is characteristic of the \textbf{Jammed Gear} regime.
In the \emph{cp} phase, the relative rotation directions vary markedly with the spatial relationship between the linkers.
As discussed above, the diagonal and central linkers are in close proximity, and their co-rotation events exhibit a pronounced preference for counter-rotation, as shown in Figure~\ref{result:gears_4}a.
The closest neighbours are counter-rotating, resembling a pair of interlocked or engaged gears, and the more distant in-plane neighbours exhibit a preference for rotation in the same direction as the central linker.
This behaviour, together with the pronounced counter-rotation of the close diagonal neighbours, conceptually resembles a three-gear arrangement in which the rotation sequence follows ($+$) ($-$) ($+$), i.e.\ the \textbf{First Gear} regime.
The First Gear pattern holds for the adjacent layers both above and below (Figure~\ref{result:gears_4}a); however, the directional preference decreases markedly beyond the first diagonal neighbour.
Thus, although each linker is coupled to several neighbours, the directional coupling along any one pathway does not propagate beyond three consecutive linkers.

In the \emph{op} phase (Figure~\ref{result:gears_4}b), the diagonal neighbours move away from the central linker, and the in-plane neighbours become the closest interaction partners (Figure~\ref{result:gears_3}b).
The central and in-plane linkers preferentially counter-rotate, indicating that they are interlocked. In contrast to the \emph{cp} phase, no preferred rotation direction is observed for the more distant linkers (Figure~\ref{result:gears_4}b).
Rotational coupling is therefore restricted primarily to pairs of in-plane linkers, corresponding to the \textbf{Second Gear} regime.

The in-plane neighbours move farther apart in the \emph{v-op} phase.
No preferential relative rotation direction remains, indicating the absence of rotational coupling between any of the linkers.
The linkers instead rotate freely and independently, resembling disengaged gears, i.e.\ the \textbf{Neutral Gear} regime.

\begin{figure}
\begin{center}
\includegraphics{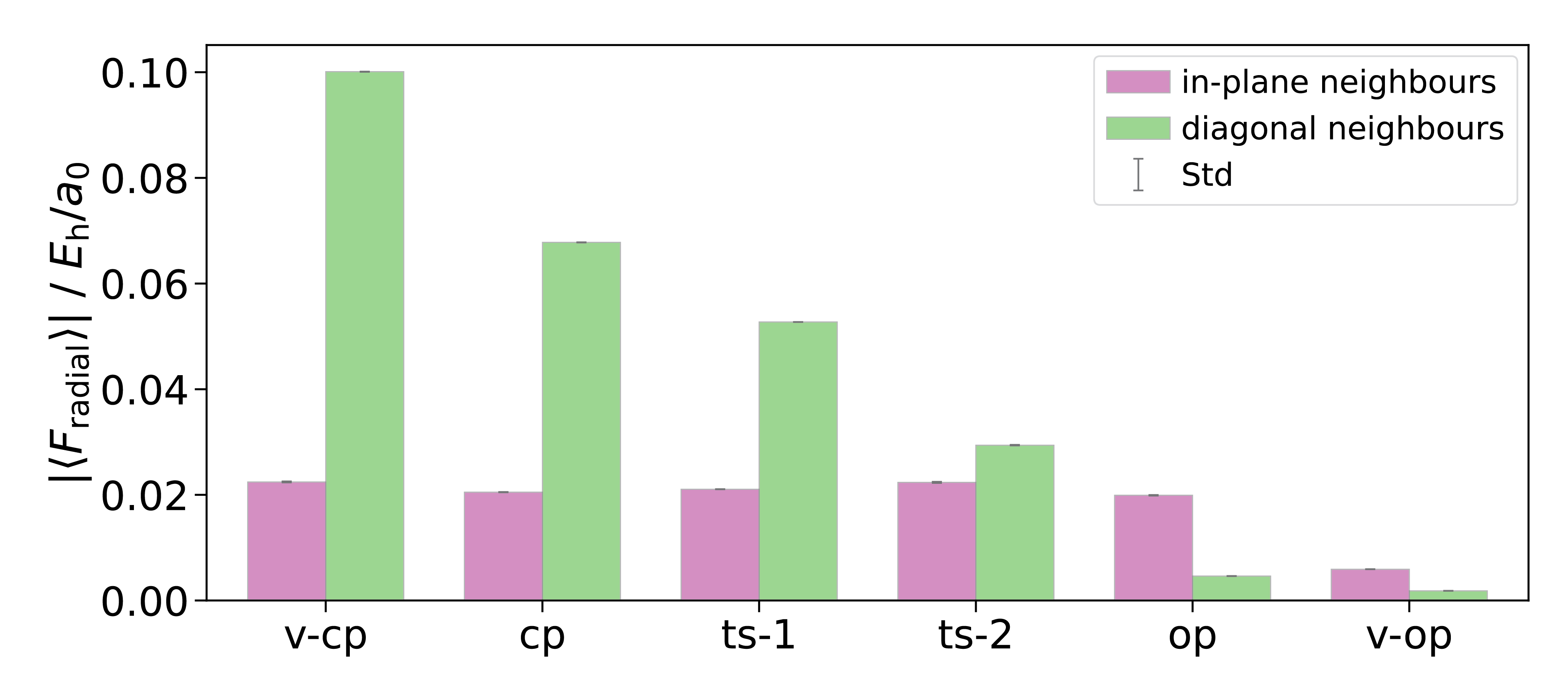}
\caption{Radial force transfer between neighbouring linkers across the different phases. The calculation is detailed in Section~S4 of the Supplementary Information. The error bars show the standard deviation between seeds.}
\label{result:gears_5}
\end{center}
\end{figure}

To move beyond a purely statistical description of linker co-rotation, the direct mechanical interactions between neighbouring linkers are also examined.
The pairwise non-bonded forces are projected onto the linker centre-to-centre direction, yielding the signed radial force transfer $F_{\mathrm{radial}}$ (where positive values indicate repulsive forces).
Figure~\ref{result:gears_5} shows the magnitude of the mean signed radial force, \(\left|\langle F_{\mathrm{radial}}\rangle\right|\), for the different phases.

In the contracted phases, the close diagonal neighbours exert substantially greater forces than the in-plane neighbours, with this difference persisting until the \emph{op} phase.
The force transferred between in-plane neighbours remains approximately constant across the phases before decreasing in the \emph{v-op} phase.
This reversal in the dominant radial force transfer from diagonal neighbours in the \emph{cp} phase to in-plane neighbours in the \emph{op} phase is consistent with the corresponding change in rotational interlocking (Figure~\ref{result:gears_4}).
Accordingly, rotational interlocking occurs primarily between diagonal neighbours in the \emph{cp} phase and between in-plane neighbours in the \emph{op} phase, distinguishing the First Gear and Second Gear regimes, respectively.

In summary, the rotational behaviour progresses from the Jammed Gear regime in the \emph{v-cp} phase, through three-linker coupling in the First Gear regime of the \emph{cp} phase and pairwise interlocking in the Second Gear regime of the \emph{op} phase, to the free and independent rotation of the Neutral Gear regime in the \emph{v-op} phase.
Considering distances, co-rotation events, and linker force transfer, we have established distinct volume-dependent linker rotation regimes.

\section*{Conclusions}
This work establishes a phase-resolved picture of linker rotation in NU-2002.
Bias-deletion metadynamics identifies the characteristic framework phases in agreement with experiment, while molecular dynamics simulations reveal their distinct rotational behaviour.
As the pore volume increases, a larger fraction of linkers rotates, yet the mean speed of the rotating linkers decreases.
The free energy barriers suppress slow rotations, causing the framework to act as an intrinsic low-speed rotational filter.
The exception is \emph{ts-2}, where a smoother, more continuous free energy landscape leads to faster rotations despite additional local barriers.
This highlights two distinct mechanisms by which the free energy landscape shapes rotation: a higher barrier selectively filters lower rotation speeds, whereas a more continuous landscape facilitates faster rotations, even at comparable barrier heights.
These insights can be leveraged to design framework materials with tailored rotational dynamics.

Inter-linker distances, radial force transfer, and co-rotation statistics reveal four distinct mechanical regimes.
In the Jammed Gear regime of the \emph{v-cp} phase, steric confinement suppresses collective rotation and permits only rare individual events.
In the First Gear regime of the \emph{cp} phase, close diagonal neighbours become rotationally interlocked, and the coupling extends across three linkers.
In the Second Gear regime of the \emph{op} phase, the dominant interaction shifts to the in-plane neighbours, producing pairwise rotational interlocking.
Finally, in the Neutral Gear regime of the \emph{v-op} phase, the increased linker separation removes directional coupling and enables independent rotation.

{NU-2002} therefore does not simply rotate more freely as its pores expand.
Instead, framework breathing switches the material between jammed, collectively geared, pairwise geared, and independent rotational regimes.
This gearbox-like response provides a mechanistic basis for controlling internal dynamics through pore geometry and a step towards the design of molecular machines.
Future work will determine how rotational energy propagates and dissipates through the framework and how these processes depend on linker geometry, metal-node connectivity (`suspension') and phonon modes.
We hope these approaches will establish the relationship between molecular rotation and macroscopic material properties.

\section*{Author contributions}
K.~S. contributed to conceptualization, investigation, formal analysis, and writing of the original draft. J.~D.~E. contributed to conceptualization, supervision, and writing through review and editing.

\section*{Conflicts of interest}
There are no conflicts to declare.

\section*{Data availability}
The data and analysis scripts supporting this study are available on Zenodo at \url{https://doi.org/10.5281/zenodo.23615975}.

\section*{Supplementary Information}
The Supplementary Information contains the individual repetitions and convergence analysis of the volume biased BDMTD simulations, the definition of the linker geometry and neighbourhood, the rotation event detection, details of the radial force transfer calculation, and the complete force field parameters.

\section*{Acknowledgements}
J.D.E. is the recipient of an Australian Research Council Discovery Early Career Award (project number DE220100163) funded by the Australian Government. We thank the Phoenix HPC service at Adelaide University for providing high-performance computing resources. This research was supported by the Australian Government’s National Collaborative Research Infrastructure Strategy (NCRIS), with access to computational resources provided by the Pawsey Supercomputing Research Centre through the National Computational Merit Allocation Scheme.
\bibliographystyle{rsc}
\bibliography{main}

\end{document}